\documentclass[11pt]{article}
\global\arraycolsep=1pt
\usepackage{subfig}

\usepackage{amsmath,amssymb,graphicx}
\usepackage{hyperref}
\usepackage{multicol,color,longtable}
\definecolor{darkred}{rgb}{0.65,0.15,0}
\hypersetup{pdfborder={0 0 0},colorlinks=true,urlcolor=blue,citecolor=blue,linkcolor=darkred,linktocpage=true}
\usepackage{nicefrac}

\usepackage[shadow,textwidth=2.7cm]{todonotes}
\usepackage{ifthen}
\reversemarginpar

\usepackage{cite}

\usepackage{tikz}
\usetikzlibrary{arrows}
\usepackage{pgfplots}

\usepackage{amsmath}
\usepackage{amsthm}
\usepackage{amssymb}
\usepackage{mathrsfs}
\usepackage[Symbol]{upgreek}
\usepackage{dsfont}
\usepackage[vcentermath]{youngtab}
\usepackage{textcomp}

\usepackage{latexsym}
\usepackage{graphicx}

\numberwithin{equation}{section}
\newcommand{\be}{\begin{equation}}
\newcommand{\ee}{\end{equation}}
\def\bea{\begin{eqnarray}}\def\eea{\end{eqnarray}}
\newcommand{\CR}{\nonumber \\*}

\newcommand{\lP}{\ell_{\scalebox{0.6}{P}}}

\newcommand{\gra}[2]{{\scriptscriptstyle (#1 , #2 )}}

\begin{document}

\begin{flushright} CPHT-RR001-012025 \end{flushright} 
 \vspace{8mm}

\begin{center}

{\LARGE \bf \sc Bounds on the next-to-leading Wilson coefficient in maximal supergravity}

\vspace{6mm}
\normalsize
{\large  Guillaume Bossard and Adrien Loty}

\vspace{10mm}

{\it Centre de Physique Th\'eorique, CNRS,  Institut Polytechnique de Paris\\
91128 Palaiseau cedex, France \footnote{email: guillaume.bossard@polytechnique.edu,  adrien.loty@polytechnique.edu}}
\vskip 1 em
\vspace{20mm}

\hrule

\vspace{5mm}

 \begin{tabular}{p{14cm}}
One expects type II string theory to be the unique ultraviolet completion of maximal supergravity. Motivated by the recent computation of the unitary bound on the leading Wilson coefficient within the S-matrix bootstrap, we analyse the minimum of the next-to-leading  Wilson coefficient in string theory in seven spacetime dimensions. We find that the next-to-leading Wilson coefficient is minimum at the same point as the leading Wilson coefficient. 
\end{tabular}

\vspace{5mm}
\hrule
\end{center}

\vspace{5mm} 

\thispagestyle{empty}

\newpage

\setcounter{page}{1}

\setcounter{tocdepth}{2}
%\tableofcontents

\section{Introduction and summary of results}
It has been argued that all theories of quantum gravity with spacetime supersymmetry can be realised in superstring theory 
\cite{Adams:2010zy,Kim:2019ths,Kim:2019vuc,Bedroya:2021fbu,Friedrich:2025gvs}. The arguments supporting this proposal are mainly based on the two-derivative low energy effective theory, and there is no proof that the complete string theory spectrum must be reproduced at high energy. The first step would  be to prove that type II string theory on a torus is the unique ultraviolet completion of maximal supergravity in  $D$-dimensional Minkowski spacetime. 

\vskip 2mm

The low-energy effective action is highly constrained by supersymmetry and U-duality, and the exact non-perturbative Wilson coefficients have been determined up to fourteen derivatives \cite{Gross:1986mw,Green:2008uj,Green:2008bf,Green:1981yb,Green:1997tv,Green:1997di,Berkovits:1997pj,Pioline:1998mn,Green:1998by,Obers:1999um,Green:1999pv,Kazhdan:2001nx,Basu:2008cf,Green:2005ba,Pioline:2010kb,Green:2011vz,Bossard:2014lra,Bossard:2014aea,Gustafsson:2014iva,Bossard:2015uga,Gourevitch:2019knu}. It was proposed in \cite{Guerrieri:2021ivu} to test the uniqueness of maximal supergravity through the unitary bound on the leading Wilson coefficient within the S-matrix bootstrap approach \cite{Paulos:2016but,Paulos:2017fhb,EliasMiro:2019kyf}. A lower bound on the leading Wilson coefficient has been computed in \cite{Guerrieri:2021ivu,Guerrieri:2022sod} in dimensions $D\ge 9$. This lower bound is very close to the minimum of the Wilson coefficient in string theory, but is not expected to be sharp because the two-to-two S-matrix ansatz used in \cite{Guerrieri:2021ivu,Guerrieri:2022sod} is not constrained to match the loop corrections in supergravity, as e.g. in  \cite{Guerrieri:2020bto}, and because they do not take into account particle production \cite{Antunes:2023irg}.

\vskip 2mm

This motivated the analysis of the minima of the non-perturbative Wilson coefficients in string theory. We analysed the leading Wilson coefficient in dimension $D\ge 6$ and the next-to-leading Wilson coefficient in $D= 8$ in \cite{Bossard:2023bhv}. In this paper we analyse  the next-to-leading Wilson coefficient in $D=7$ dimensions.

\vskip 4mm

%\section{Notations and summary of results}

We first introduce our notations, which are the same as in \cite{Bossard:2023bhv}. The non-perturbative four-graviton amplitude in type II string theory  on $\mathds{R}^{1,6}\times T^3$ factorises into
\be \mathcal{M}_{4} = - i \frac{\kappa_{\scalebox{0.6}{$7$}}^2}{2^{10}} t_8 t_8 \prod_{a=1}^4 R(k_a,\epsilon_a) \mathcal{A}(s,t,u,\varphi)  \label{ScalarFunction} \ee
where $k_a$ and $\epsilon_a$ are respectively the momenta and polarisations of the four gravitons, while $\mathcal{A}(s,t,u,\varphi)$ is invariant under permutations of the three Mandelstam variables and is a function of the scalar fields expectation values  $\varphi$ that parametrise the double coset moduli space  $\mathcal{M}_5 = SO(5) \backslash SL(5,\mathds{R})/ SL(5,\mathds{Z})$. We define the Planck length in seven spacetime dimensions as
\be \kappa_{\scalebox{0.6}{$7$}}^2 = \frac12 (2\pi)^{4} \lP^{5} \; , \ee
and the $7$-dimensional effective string coupling $g_{\scalebox{0.6}{$7$}} = e^{\phi} /\sqrt{\upsilon_3}$ in terms of the ten-dimensional dilaton $\phi$ and the volume Vol$(T^3) = (2\pi \sqrt{\alpha^\prime} )^3 \upsilon_3 $ of the 3-torus in string length, such that 
\be \alpha^\prime = g_{\scalebox{0.6}{$7$}}^{- \frac{4}{5}} \lP^2 \; . \label{EinsteinString}\ee
In the low energy limit, the expansion of $\mathcal{A}(s,t,u,\varphi)$ gives \cite{Green:1999pv,Green:2008uj}
\begin{multline} \mathcal{A}(s,t,u,\varphi)  = \frac{64}{s t u} + 32 (2\pi)^{4} \lP^{5} \bigl( I_4^{(1)}(s,t) + I_4^{(1)}(t,u) + I_4^{(1)}(u,s)\bigr) + \lP^6 \mathcal{E}_\gra{0}{0}(\varphi)\\
+16(2\pi)^{8}\lP^{10}\biggl(s^2(I_{4}^{(2)}(s,t)+I_{4}^{(2)}(s,u))
+t^2(I_{4}^{(2)}(t,u)+I_{4}^{(2)}(t,s))+u^2(I_{4}^{(2)}(u,s)+I_{4}^{(2)}(u,t))\biggr)\\
+ \frac{ \lP^{10}}{16}  \mathcal{E}_\gra{1}{0}(\varphi)(s^2+t^2+u^2)   + o( \lP^{10})  \label{LowEnergyExpansion} \end{multline} 
where $I_4^{(1)}$ represents the scalar box diagram integral and $I_4^{(2)}$ is the two-loop scalar integral that we define in dimensional regularisation as
\begin{multline} I_{4,\epsilon}^{(2)}(s,t) =  \lP^{-4\epsilon} \int \frac{d^{7-2\epsilon}p d^{7-2\epsilon} q}{(2\pi)^{14-4\epsilon}}  \Biggl(  \frac{1}{p^2 (p{-}k_1)^2 (p{-}k_1{-}k_2)^2(p{+}q)^2 q^2(q{-}k_4)^2 (q{-}k_3{-}k_4)^2} \\
+ \frac{1}{p^2 (p{-}k_1)^2(p{-}k_1{-}k_2)^2 (p{+}q)^2 (p{+}q{+}k_3)^2 q^2(q{-}k_4)^2}  \Biggr)\; .  \end{multline}
This integral has a pole at $\epsilon=0$ associated to the 2-loop logarithmic divergence in supergravity \cite{Bern:2008pv}. In string theory the full amplitude is finite, and this divergence is absorbed in the definition of the next-to-leading Wilson coefficient $\lP^{10} \mathcal{E}_\gra{1}{0}(\varphi)$. We define the renormalised two-loop diagram as
\be  I_{4}^{(2)}(s,t) = \lim_{\epsilon\rightarrow 0} \Biggl( I_{4,\epsilon}^{(2)}(s,t)  - \frac{1}{(4\pi)^8} \Bigl( \frac{\pi^2}{3\epsilon} + 2\pi^2 \log(2\pi) - 4 \zeta'(2) \Bigr) \Biggr) \; ,  \label{RenormaliseI4} \ee
such that the Wilson coefficient is of `uniform transcendentality'. We will determine this renormalisation in Section \ref{The superstring two-loop integral}, by analysing the low energy expansion of the two-loop superstring amplitude.\footnote{In principle one may want to define the renormalised Feynman integral $I_{4}^{(2)}(s,t)$ to be of uniform transcendentality and to define the Wilson coefficient accordingly. But for this one would first need to compute the integral in seven dimensions. Note that there is no reason to expect the total string amplitude be of uniform transcendentality, since this is not the case at one-loop in ten dimensions \cite{DHoker:2019blr,Claasen:2024ssh}.}

 The next-to-leading Wilson coefficient is given in terms of regularised  Eisenstein series for the U-duality group $SL(5)$ as \cite{Green:2010wi}
\be
 \mathcal{E}_\gra{1}{0}(\varphi) =\zeta(5)\widehat{E}^{\scalebox{0.65}{$SL(5)$}}_{\frac52\Lambda_1}(H)+\frac{\pi}{15}\zeta(5)\widehat{E}^{\scalebox{0.65}{$SL(5)$}}_{\frac52\Lambda_3}(H)
 \label{Wilson coeff}
\ee
where $H(\varphi)$ is a symmetric positive definite unimodular $5\times5$ matrix parametrising the coset $SO(5) \backslash SL(5,\mathds{R})$, $\Lambda_{1}$ and $\Lambda_3$ are fundamental weights of $SL(5)$ in the Bourbaki convention for labelling simple roots and $\zeta(s)$ is the Riemann zeta function. At a regular value of the parameter $s$ the maximal parabolic $SL(5)$ Eisenstein series can be defined by analytic continuation of the  sums
\bea &&E^{\scalebox{0.65}{$SL(5)$}}_{s\Lambda_1}(H) = \frac{1}{2\zeta(2s)}\sum_{n\in \mathds{Z}^5 \smallsetminus \{0\} } \frac{1}{H[n]^s}  \label{Eisenstein5} \;,\\
&&E^{\scalebox{0.65}{$SL(5)$}}_{s\Lambda_3}(H) = \frac{1}{2\zeta(2s)}\sum_{\substack{q\in \bigwedge^{\!2}\! \mathds{Z}^5 \smallsetminus \{0\} \\q\wedge q=0}} \frac{1}{(H^{-1}\wedge H^{-1}[q])^s}  \; , 
\label{Eisenstein5 lambda2}
\eea
where we used the notations $H[n]=n^\intercal Hn$ for $n\in \mathds{Z}^5$ and $H^{-1}\wedge H^{-1}[n_1\wedge n_2]$ is a shorthand for $(n_1^\intercal H^{-1}n_1)(n_2^\intercal H^{-1}n_2)-(n_1^\intercal H^{-1}n_2)^2$ with $n_1,n_2\in \mathds{Z}^5$ and not collinear. The definition of the renormalised Eisenstein series $\widehat{E}^{\scalebox{0.55}{$SL(5)$}}_{\scalebox{0.6}{$\frac{5}{2}\Lambda_{1}$}}$ and $\widehat{E}^{\scalebox{0.55}{$SL(5)$}}_{\scalebox{0.6}{$\frac{5}{2}\Lambda_{3}$}}$ is chosen such that there is no additional transcendental constant and the logarithm of the string coupling constant and the torus radii in string units are defined to vanish at the self-dual points. The precise definition is given  in Section \ref{Numerical}.

We have analysed the minimum of the Eisenstein series $E^{\scalebox{0.65}{$SL(5)$}}_{s \Lambda_1}$ in \cite{Bossard:2023bhv}, where we provided strong evidence for the global minimum of the function to be at $H = H_{D_5}$, where $H$ is conformal to the $D_5$ lattice bilinear form,  for all values of $s>\frac{5}{4}$. This result was already proved for large $s$, and more generally, $E^{\scalebox{0.65}{$SL(N)$}}_{s \Lambda_1}$ admits a global minimum for large enough $s$ where  $H$ defines the metric of the densest lattice sphere packing \cite{MinimaLatticePacking}. 

The Eisenstein series $E^{\scalebox{0.65}{$SL(5)$}}_{s \Lambda_1}$ admits only non-zero Fourier coefficients for unipotent characters in the minimal co-adjoint orbit. One says accordingly that it is attached to a minimal representation, and defines therefore the simplest class of automorphic functions on $\mathcal{M}_5$. It is therefore expected that its extrema are special points of the symmetric space. The symmetric space admits a stratification determined by the stabilizer subgroup $G^H(\mathds{Z}) \subset SL(5,\mathds{Z})$ of $H$. A point $H$ is in a stratum of dimension $n\le 14$ if the $G^H(\mathds{Z})$-invariant subspace of $T \mathcal{M}_5|_H$ is of dimension $n$. The main stratum of dimension $n=14$ is the open subset of the fundamental domain. The dimension zero stratum is the set of isolated points that we call symmetric points. This stratification can easily be described in the simpler case of the upper complex half plane with the action of $GL(2,\mathds{Z})$. The main stratum is then the open fondamental domain where $0< {\rm Re}[\tau] < \frac12$ and $|\tau|>1$. The dimension 1 stratum includes the line ${\rm Re}[\tau] = \frac12$ invariant under $\tau\rightarrow 1-\bar \tau$, as well as the arc $|\tau|=1$, invariant under $\tau\rightarrow \frac{1}{\bar \tau}$. The dimension 0 stratum is the union of the two symmetric points, $\tau =  e^{i \frac{\pi}{3}}$ invariant under S$_3$ and $\tau = i $ invariant under $\mathds{Z}_2\times \mathds{Z}_2$.

By construction, symmetric points are extrema of any $SL(5,\mathds{Z})$ invariant function on the symmetric space $SO(5) \backslash SL(5,\mathds{R})$. In  \cite{Bossard:2023bhv} we provided strong evidence for the conjecture that all local minima of $E^{\scalebox{0.65}{$SL(5)$}}_{s \Lambda_1}$ are located on symmetric points. Assuming this is the case, finding the global minimum amounts to classifying the symmetric points and checking for which one $E^{\scalebox{0.65}{$SL(5)$}}_{s \Lambda_1}$ is the smallest. One can prove generally that reducible symmetric points, for which $H$ defines a reducible lattice, are always saddle points. One can therefore concentrate on irreducible symmetric points and we proved that $H_{D_5}$ gives the minimum value of the Eisenstein series.

In order to analyse the next-to-leading Wilson coefficient, we generalise this analysis to the Eisenstein series ${E}^{\scalebox{0.65}{$SL(5)$}}_{s \Lambda_2}$ in this paper. It admits non-zero Fourier coefficients for unipotent characters in the minimal and the next-to-minimal co-adjoint orbits. As such one says that it is attached to a next-to-minimal representation, and still defines a very simple class of automorphic functions on $\mathcal{M}_5$. We analyse this function on a dimension-two surface inside a dimension-three stratum of points stabilised by S$_4\subset SL(5,\mathds{Z})$, and find as for the minimal Eisenstein series that the local minima (located on the surface) are at the irreducible symmetric points.

In this paper we support the conjecture that the global minimum of $\widehat{E}^{\scalebox{0.65}{$SL(5)$}}_{\frac{5}{2} \Lambda_3}$ is at $H_{D_5^*}$, while
\vspace{-1mm}

\noindent the global minimum of $\widehat{E}^{\scalebox{0.65}{$SL(5)$}}_{\frac{5}{2} \Lambda_1}$ is at $H_{D_5}$. It is therefore natural to expect the next-to-leading Wilson coefficient minimum to be on the stratum of dimension 1 inside $\mathcal{M}_5$ linking these two points. With these hypothesis and with the help of numerical analysis we find that the global minimum is in fact at $H_{D_5}$ as for the leading Wilson coefficient. Indeed by plotting the behavior of both functions on the dimension 1 stratum between $H_{D_5}$ and $H_{D_5^*}$ we can see graphically that their variations are nearly identical in the sense that their derivatives nearly always have the same sign. Hence the next-to-leading Wilson coefficient which is obtained by a positive linear combination of the two also has the same behavior and in particular the same local minima. We then checked that the value at $H_{D_5}$ is the smallest. We evaluate numerically the value of the next-to-leading Wilson coefficient with the renormalisation prescription \eqref{RenormaliseI4} as
\be
 \mathcal{E}_\gra{1}{0}(\varphi_{D_5}) \approx26.2315\;.
\ee

\section{Numerical approximations}
\label{Numerical} 
The main purpose of this paper is to find the lower bound of the next-to-leading Wilson coefficient of maximally supersymmetric string theory in dimension $D=7$. We shall first discuss more generally the Eisenstein series of $SL(N)$. In \cite{Bossard:2023bhv} we analysed the minimal Eisenstein series $E^{\scalebox{0.65}{$SL(N)$}}_{s\Lambda_{1}}$ and here we will generalise this analysis to the next-to-minimal Eisenstein series $E^{\scalebox{0.65}{$SL(N)$}}_{s\Lambda_{2}}$. To derive a numerical approximation of these functions in the interior of the moduli space for values of the parameter $s$ close to $\frac{N}{2}$, we need to compute the Fourier expansion of these functions.\footnote{Only for Re$[s]\gg N$ is it more convenient to use the definitions \eqref{Eisenstein5} and \eqref{Eisenstein5 lambda2}.} To write this expansion it is convenient to use an abelian enumeration of roots according to \cite{Gourevitch:2019knu}, which we take to be defined iteratively by the inclusion of the abelian parabolic 
\be
P_1=(GL(1)\times SL(N{-}1))\ltimes\mathds{R}^{N-1}\subset SL(N)\;,
\label{P_1}
\ee
 associated to the fundamental weight $\Lambda_1$ of $SL(N)$, for all $N$. We take the Iwasawa decomposition compatible with the enumeration of roots such that $\mathcal{V}_N$ defines the coset representative in the Borel subgroup $B_N$
 \be 
 \mathcal{V}_N = \begin{pmatrix}
r_{\scalebox{0.6}{$N$}}^{\frac{1-N}{2}} & r_{\scalebox{0.6}{$N$}}^{\frac{1-N}{2}} x_{\scalebox{0.6}{$N$}}^\intercal \\
0 & r_{\scalebox{0.6}{$N$}}^{\frac{1}{2}} \mathcal{V}_{N-1}
\end{pmatrix} \;,
 \ee
with $r_{\scalebox{0.6}{$N$}}>0 $, $\mathcal{V}_{N-1}\in B_{N-1}$ and $x_{\scalebox{0.6}{$N$}}\in\mathds{R}^{N-1}$ for all $N$. We will identify $SO(N)\backslash SL(N)$ with the space $\mathcal{S}_N$ of symmetric positive definite $N\times N$ matrices of unit determinant $H_N=\mathcal{V}_N^\intercal \mathcal{V}_N$, with  
 \be
H_N=\begin{pmatrix}
1& 0 \\
x_{\scalebox{0.6}{$N$}} &\mathds{1}
\end{pmatrix} \begin{pmatrix}
r_{\scalebox{0.6}{$N$}}^{1-N} & 0\\
0 & r_{\scalebox{0.6}{$N$}} H_{N-1}
\end{pmatrix}\begin{pmatrix}
1& x_{\scalebox{0.6}{$N$}}^\intercal\\
0 &\mathds{1}
\end{pmatrix}\; . 
\label{parabolic decomp}
\ee
For short we shall drop the index $N$ of $r_{\scalebox{0.6}{$N$}}$ and $x_{\scalebox{0.6}{$N$}}$ in the following. Using the Poisson summation formula, one straightforwardly works out the Fourier expansion of the minimal Eisenstein series as 
\bea
E^{\scalebox{0.65}{$SL(N)$}}_{s\Lambda_{1}}(H_N)=&&r^{(N-1)s}+\frac{\xi(2s-1)}{\xi(2s)}r^{\frac{N}{2}-s}E^{\scalebox{0.65}{$SL(N{-}1)$}}_{(s-\frac{1}{2})\Lambda_{1}}(H_{N-1})\nonumber\\
&&+\frac{2}{\xi(2s)}\sum_{\Gamma\in\mathds{Z}^{N-1}}^\prime\sigma_{2s-1}(\Gamma)\frac{r^{(\frac{N}{2}-1)s+\frac{N}{4}}}{\lvert Z(\Gamma)\lvert^{s-\frac{1}{2}}}K_{s-\frac{1}{2}}(2\pi r^\frac{N}{2}\lvert Z(\Gamma)\lvert)\cos(2\pi(\Gamma,x))\;,
\label{1 Fourier series}
\eea
where ${}\prime$ on the sum indicates that $\Gamma=0$ is omitted, $\lvert Z(\Gamma)\lvert^2=\Gamma^\intercal H_{N-1}\Gamma$, $(\Gamma,x)=\Gamma^\intercal x$, and $\sigma_s(\Gamma)$ is the sum of the divisors of the greatest common divisor of $\Gamma$ to the power $s$.  For $N=5$ the scalar $r_5^{\frac{5}{2}} = \frac{1}{g_{\scalebox{0.6}{$7$}}}$ is the inverse string coupling and $H_4$ decomposes into $r_4 H_3 = G^{\scalebox{0.6}{IIB}}$ and $x_4=B^{\scalebox{0.6}{IIB}}$ which are respectively the type IIB metric  in string frame and the Kalb-Ramond field on $T^3$. The Fourier mode $\Gamma\in \mathds{Z}^4$ is the Dp-brane instanton charge for Euclidean branes wrapping $T^3$. Then $\lvert Z(\Gamma)\lvert$ is the central charge of the instanton solution and $2\pi/g_{\scalebox{0.6}{$7$}} |Z(\Gamma)| + 2\pi i (x_5,\Gamma)$ its Euclidean action. We will refer to $\lvert Z(\Gamma)\lvert$ as the central charge for all $N$ by abuse of language.

 The Eisenstein series can be seen to admit an analytic continuation in $s\in \mathds{C}- \{ \frac{N}{2}\}$, using the properties of the Bessel function of the second kind $K_s$ and the completed zeta function $\xi(s)=\pi^{-s/2}\Gamma(s/2)\zeta(s) = \xi(1-s)$. The anti-fundamental Eisenstein series satisfies the functional relation 
\be E^{\scalebox{0.65}{$SL(N)$}}_{s\Lambda_{N-1}}(H_N) = E^{\scalebox{0.65}{$SL(N)$}}_{s\Lambda_{1}}(H_N^{-1}) = \frac{\xi(2s+1-N)}{\xi(2s)} E^{\scalebox{0.65}{$SL(N)$}}_{(\frac{N}{2}-s)\Lambda_{1}}(H_N)\; , \label{MinLFunc} \ee
as one can check using 
\bea
E^{\scalebox{0.65}{$SL(N)$}}_{s\Lambda_{N-1}}(H_N)=&&r^{s}E^{\scalebox{0.65}{$SL(N{-}1)$}}_{s\Lambda_{N-2}}(H_{N-1})+\frac{\xi(2s-N+1)}{\xi(2s)}r^{(N-1)(\frac{N}{2}-s)}\nonumber\\
&&+\frac{2}{\xi(2s)}\sum_{\Gamma\in\mathds{Z}^{N-1}}^\prime\sigma_{N-1-2s}(\Gamma)\frac{r^{(1-\frac{N}{2})s+\frac{N(N-1)}{4}}}{\lvert Z(\Gamma)\lvert^{\frac{N-1}{2}-s}}K_{s-\frac{N-1}{2}}(2\pi r^\frac{N}{2}\lvert Z(\Gamma)\lvert)\cos(2\pi(\Gamma,x))\; .\nonumber\\
\label{N-1 Fourier series}
\eea

One computes similarly for the next-to-minimal Eisenstein series \footnote{This computation can be done following the same steps as in Appendix B of \cite{Bossard:2015oxa}. }
\bea
&&E^{\scalebox{0.65}{$SL(N)$}}_{s\Lambda_{2}}(H_N)=r^{(N-2)s}E^{\scalebox{0.65}{$SL(N{-}1)$}}_{s\Lambda_{1}}(H_{N-1})+\frac{\xi(2s-2)}{\xi(2s)}r^{N-2s}E^{\scalebox{0.65}{$SL(N{-}1)$}}_{(s-\frac{1}{2})\Lambda_{2}}(H_{N-1})\nonumber\\
&&+\frac{2}{\xi(2s)}\sum_{\Gamma\in\mathds{Z}^{N-1}}^\prime\frac{\sigma_{2-2s}(\Gamma)}{\text{gcd}(\Gamma)^{\frac{N-3}{N-2}(1-2s)}}E^{\scalebox{0.65}{$SL(N{-}2)$}}_{(s-\frac{1}{2})\Lambda_{1}}(H_\Gamma)\frac{r^{(\frac{N}{2}-2)s+\frac{N}{2}}}{\lvert Z(\Gamma)\lvert^\frac{(N-4)s+1}{N-2}}K_{s-1}(2\pi r^\frac{N}{2}\lvert Z(\Gamma)\lvert)\cos(2\pi(\Gamma,x))\;.\nonumber\\
\label{N-2 Fourier series}
\eea
where $H_\Gamma\in\mathcal{S}_{N-2}$ is the representative of $SO(N{-}2)\backslash SL(N{-}2)$ in the Levi stabilizer $SL(N{-}2)\subset SL(N{-}1)$ of $\Gamma$. The next-to-minimal Eisenstein series admits an analytic continuation in $s\in \mathds{C}- \{ \frac{N}{2},\frac{N-1}{2}\}$, and satisfies the functional relation 
\be E^{\scalebox{0.65}{$SL(N)$}}_{s\Lambda_{N-2}}(H_N) = E^{\scalebox{0.65}{$SL(N)$}}_{s\Lambda_{2}}(H_N^{-1}) = \frac{\xi(2s+1-N)\xi(2s+2-N)}{\xi(2s)\xi(2s-1)} E^{\scalebox{0.65}{$SL(N)$}}_{(\frac{N}{2}-s)\Lambda_{2}}(H_N)\; . \label{NextLFunc}\ee

To evaluate this function we need to define $H_\Gamma$ explicitly in the Iwasawa coordinates. Let us explain this is detail. Any $\Gamma \in \mathds{Z}^{N-1}$ is in the $SL(N{-}1,\mathds{Z})$ orbit of a canonical element $\Gamma_0=\text{gcd}(\Gamma)\left(1,0,\dots,0\right)^\intercal$. We write  $\gamma\in SL(N{-}1,\mathds{Z})$ such that $\Gamma=\gamma\Gamma_0$. The parabolic subgroup $P_1\subset SL(N{-}1)$ is by definition the (conformal) stabilizer of $\Gamma_0$ so that $\mathcal{V}_{N{-}1} \Gamma_0 = r_{N{-}1}^{\frac{2{-}N}{2}}\Gamma_0 $. To define the action of $\gamma$ on the symmetric space, one takes the Iwasawa decomposition $\gamma^\intercal H_{N-1}\gamma=b_\Gamma^\intercal b_\Gamma$ for $b_\Gamma \in B_{N{-}1}$. By definition of the Borel subgroup, $b_\Gamma \Gamma_0 =y\Gamma_0 $ for $y>0$ the square root of the first component of the matrix $\gamma^\intercal H_{N-1}\gamma$, i.e. $y^2 =\left(1,0,\dots,0\right) \gamma^\intercal H_{N-1}\gamma \left(1,0,\dots,0\right)^\intercal$. This determines $H_\Gamma = \mathcal{V}^\intercal_{\Gamma}\mathcal{V}_{\Gamma}$ with $\mathcal{V}_{\Gamma}$ defined such that \footnote{We can define all the elements $k_\Gamma b_\Gamma \gamma^{-1}\in SL(N{-}1)$ that stabilise $\Gamma$ up to a rescaling such that 
\be k_\Gamma b_\Gamma \gamma^{-1} \Gamma = \frac{{\rm gcd}(\Gamma)}{|\Gamma|} y \Gamma\; , \ee
where $|\Gamma|$ is the Euclidean norm of $\Gamma$ and $k_\Gamma \in SO(N{-}1)$ is determined such that $k_\Gamma\Gamma_0=\frac{{\rm gcd}(\Gamma)}{|\Gamma|} \Gamma$. The existence of $k_\Gamma$ is ensured by the property that the action of $SO(N{-}1)$ is transitive on the spheres of fixed radius.}
\be 
b_\Gamma = \begin{pmatrix}
y  &  y  x_{\Gamma}^\intercal \\
0 & y^{\frac{1}{2-N}}  \mathcal{V}_{\Gamma}
\end{pmatrix} \; .
 \ee
 The only non-trivial part is therefore to determine $\gamma$.  Its first row is $\frac{\Gamma}{\text{gcd}(\Gamma)} $ by construction, so that we need to find $\gamma \in SL(N{-}1,\mathds{Z})$ of the form 
 \be
\gamma=\begin{pmatrix}
  &  & \\
\frac{\Gamma}{\text{gcd}(\Gamma)} &\   & *\ \\
 & &
\end{pmatrix}\;,
\ee
which amounts to solving a set of diophantine equations which are tractable for $N\le 5$.\\

With these formulas we have all the definitions required to obtain a numerical approximation of the next-to-minimal Eisenstein series $E^{\scalebox{0.65}{$SL(5)$}}_{s\Lambda_{2}}$ at a generic $s$. Using the asymptotic expansion of the Bessel function $K_s(z)\sim\sqrt{\frac{\pi}{2z}}e^{-z}$ one can see that the non-constant Fourier coefficients in \eqref{N-2 Fourier series} will be exponentially suppressed for charges $\Gamma$ with large central charge $\lvert Z(\Gamma)\lvert$. Therefore we can focus on charges with low values of $\lvert Z(\Gamma)\lvert$. The classification of charges of fixed low central charge $\lvert Z(\Gamma)\lvert$ depends on the point in moduli space. It turns out that it is numerically less expensive to just consider charges  in $\mathds{Z}^4$ with integer entries between $-n_{\scalebox{0.6}{max}}$ and $n_{\scalebox{0.6}{max}}$, although only a subset of these charges define a complete subset of charges with $\lvert Z(\Gamma)\lvert$ bounded from above by some fixed $\lvert Z_{\scalebox{0.6}{max}}\lvert$. In practice we find that $n_{\scalebox{0.6}{max}} = 5$ provides  an excellent approximation, giving the first 22 layers of charges of fixed central charge for the symmetric points we are mostly interested in. We have checked that going from $n_{\scalebox{0.6}{max}}=4$ to $n_{\scalebox{0.6}{max}}=5$ only modifies the function by a negligible amount in the deep interior of the moduli space, although this approximation may become inaccurate near the cusps where the function is very large and therefore far away from its minimum.

\vskip 5mm 

The next-to-leading Wilson coefficient is in fact defined from the finite part of Eisenstein series evaluated at the singular value $s= \frac{N}{2}$ for $N=5$. We therefore define the renormalised Eisenstein series $\widehat{E}^{\scalebox{0.65}{$SL(5)$}}_{\frac52\Lambda_1}$ and $\widehat{E}^{\scalebox{0.65}{$SL(5)$}}_{\frac52\Lambda_3}$ as the finite limits 
\bea
\widehat{E}^{\scalebox{0.65}{$SL(N)$}}_{\frac{N}{2}\Lambda_{1}}\hspace{-1mm}&=&\lim_{\epsilon\rightarrow0}\left(E^{\scalebox{0.65}{$SL(N)$}}_{(\frac{N}{2}+\epsilon)\Lambda_{1}}-\frac{\xi(1+2\epsilon)}{\xi(N+2\epsilon)}\right)\; ,\CR
\widehat{E}^{\scalebox{0.65}{$SL(N)$}}_{\frac{N}{2}\Lambda_{N-2}}\hspace{-1mm}&=&\lim_{\epsilon\rightarrow0}\left(E^{\scalebox{0.65}{$SL(N)$}}_{(\frac{N}{2}+\epsilon)\Lambda_{N-2}}-\frac{\xi(1+2\epsilon)\xi(2+2\epsilon)}{\xi(N+2\epsilon)\xi(N-1+2\epsilon)}\right)\; ,
\eea
where the subtraction is chosen as the factor in the functional relations \eqref{MinLFunc} and \eqref{NextLFunc}. This ensures that their Fourier expansions only involve the logarithm of the moduli $r_i$, and no other transcendental terms coming from the expansion of $\xi(s)$ at the pole.

To evaluate numerically the next-to-minimal Eisenstein series $\widehat{E}^{\scalebox{0.55}{$SL(5)$}}_{\scalebox{0.6}{$\frac{5}{2}\Lambda_{3}$}}(H^{-1})=\widehat{E}^{\scalebox{0.55}{$SL(5)$}}_{\scalebox{0.6}{$\frac{5}{2}\Lambda_{2}$}}(H)$ one uses the Fourier expansion of the renormalised Eisenstein series 
\bea
&&\widehat E^{\scalebox{0.65}{$SL(N)$}}_{\frac{N}{2}\Lambda_{2}}(H_N)=r^{\frac{N(N-2)}{2}}E^{\scalebox{0.65}{$SL(N{-}1)$}}_{\frac{N}{2}\Lambda_{1}}(H_{N-1})+\frac{\xi(N-2)}{\xi(N)}\widehat  E^{\scalebox{0.65}{$SL(N{-}1)$}}_{\frac{N-1}{2}\Lambda_{2}}(H_{N-1})-\frac{\xi(2)}{\xi(N-1)\xi(N)}\ln r\nonumber\\
&&+\frac{2}{\xi(N)}\sum_{\Gamma\in\mathds{Z}^{N-1}}^\prime\frac{\sigma_{N-2}(\Gamma)}{\text{gcd}(\Gamma)^{\frac{1}{N-2}}}E^{\scalebox{0.65}{$SL(N{-}2)$}}_{\frac{N-1}{2}\Lambda_{1}}(H_\Gamma)\frac{r^{\frac{N(N-2)}{4}}}{\lvert Z(\Gamma)\lvert^{\frac{N{-}2}{2}-\frac{1}{N-2}}}K_{\frac{N-2}{2}}(2\pi r^\frac{N}{2}\lvert Z(\Gamma)\lvert)\cos(2\pi(\Gamma,x))\;.\nonumber\\
\label{2 regularised Fourier series}
\eea
successively for $N\le 5$, as well as the expansion of $E^{\scalebox{0.65}{$SL(N)$}}_{\scalebox{0.6}{$\frac{N{+}1}{2}\Lambda_{1}$}}$ given in \eqref{1 Fourier series} for $N\le 4$. However, the truncated sum over the Fourier coefficients is difficult to evaluate as a function of the 14 variables parametrising $SO(5) \backslash SL(5)$, mainly because of the evaluation of $E^{\scalebox{0.65}{$SL(3)$}}_{\scalebox{0.7}{$2\Lambda_{1}$}}(H_\Gamma)$ for many Fourier coefficients, because $H_\Gamma$ is already a complicated function of the moduli for many charges. Following \cite{Bossard:2023bhv}, we shall therefore argue instead that the global minimum must be at a special point and we shall only evaluate this function numerically on a special surface inside  $SO(5) \backslash SL(5)$.

\vskip 5mm

For this purpose, it is useful to define the stabilizer group of a point $\mathcal{V}_0\in SO(N)\backslash SL(N)$ as the finite subgroup
\be 
G^{\mathcal{V}_0}(\mathds{Z})=\bigl\{\gamma\in SL(N,\mathds{Z})\ \lvert\ \gamma^\intercal \mathcal{V}^\intercal_0 \mathcal{V}_0 \gamma = \mathcal{V}^\intercal_0 \mathcal{V}_0 \bigr\} =\bigl\{\gamma\in SL(N,\mathds{Z})\ \lvert\ \mathcal{V}_0\gamma\mathcal{V}_0^{-1}\in SO(N)\bigr\}\; . 
\ee
The second definition shows that the discrete group  $G^{\mathcal{V}_0}(\mathds{Z})$ can always be represented by $SO(N)$ matrices $k_\gamma$, and is therefore finite. For a given stabilizer $G^{\mathcal{V}_0}(\mathds{Z})$, one can then define the set of points $\Sigma_{\mathcal{V}_0}$ in $SO(N)\backslash SL(N)$ that are invariant under the same group, i.e.\footnote{Here we have defined the closed set, for the open set one would simply replace $G^{\mathcal{V}_0}(\mathds{Z})\subset G^{\mathcal{V}}(\mathds{Z})$ in the definition by $G^{\mathcal{V}_0}(\mathds{Z})= G^{\mathcal{V}}(\mathds{Z})$. In this way, it is clear that this definition is independent of $\mathcal{V}_0$ for almost all points $\mathcal{V}_0\in\Sigma_{\mathcal{V}_0}$.}
\be
\Sigma_{\mathcal{V}_0}=\bigl\{\mathcal{V}\in SO(N)\backslash SL(N)\ \lvert\ G^{\mathcal{V}_0}(\mathds{Z})\subset G^{\mathcal{V}}(\mathds{Z})\bigr\}\; .
\label{symmetric stratum}
\ee
In \cite{Bossard:2023bhv} we define symmetric points as the points $\mathcal{V}_0$ in the moduli space for which $\Sigma_{\mathcal{V}_0}$ is a set of isolated points. Symmetric points are extrema of all automorphic functions. One can understand this using the Taylor expansion of an automorphic function $f(\mathcal{V})$ 
 \be
f_{\mathcal{V}_0}(p) \equiv f(\exp(p) {\mathcal{V}_0})= f(\mathcal{V}_0)+p^aD_af(\mathcal{V}_0)+\frac{1}{2}p^ap^bD_aD_bf(\mathcal{V}_0)+\mathcal{O}(p^3)\; , 
\ee
where $p\in \mathfrak{sl}(N)\ominus \mathfrak{so}(N)$ of components $p^a$ and $D_a$ is the covariant derivative on the symmetric space in tangent frame.
For all $k_\gamma = \mathcal{V}_0\gamma\mathcal{V}_0^{-1}\in SO(N)$  defined from  $\gamma \in G^{\mathcal{V}_0}(\mathds{Z})$ we have that 
\be f_{\mathcal{V}_0}(p) =  f_{\mathcal{V}_0}(k_\gamma^{-1} pk_\gamma)\ee
and at each order the polynomials in $p$ appearing in the Taylor expansion of $f$ are invariant under the action of $\mathcal{V}_0 G^{\mathcal{V}_0}(\mathds{Z})\mathcal{V}_0^{-1}\subset SO(N)$. For a symmetric point there is no linear polynomial of $p$ invariant under $G^{\mathcal{V}_0}(\mathds{Z})$, and $D_a f(\mathcal{V}_0)=0$. The Hessian matrix $D_aD_bf(\mathcal{V}_0)$ must also be an invariant polynomial of degree two, reducing the number of independent conditions to be checked to determine if a symmetric point 
extremum is a minimum or not.\\

We showed in \cite{Bossard:2023bhv} that all the symmetric points for which $H$ is a reducible matrix are necessarily saddle points. We can repeat the same proof for $E^{\scalebox{0.55}{$SL(5)$}}_{s\Lambda_{2}}$ and $E^{\scalebox{0.55}{$SL(5)$}}_{s\Lambda_{3}}$ using their maximal parabolic Fourier expansions for $s>2$. The only difference is that the Fourier coefficients now get a term proportional to $E^{\scalebox{0.55}{$SL(3)$}}_{\scalebox{0.6}{$(s{-}\frac12)\Lambda_{1}$}}(H_\Gamma)$ in $P_1$ or $E^{\scalebox{0.55}{$SL(2)$}}_{\scalebox{0.6}{$(s{-}1)\Lambda_{1}$}}(H_\Gamma)$ in $P_2$, and are strictly positive provided $s>2$, since these Eisenstein series are then absolutely convergent. The only symmetric points we are interested in to obtain local minima are therefore the irreducible symmetric matrices with a maximal symmetry group which were classified in \cite{AutoL5}. The corresponding symmetric matrices, that we write  $H_L$, are proportional to the Gram matrix of even lattices $L$. The irreducible lattices with maximal symmetry group are the root lattices $D_5$ and $A_5$, their duals, and the lattice $A_5^{+2}$ generated  by $A_5$ and the fundamental weight $\Lambda_2$ and its dual $A_5^{+3}$ that is generated by $A_5$ and the fundamental weight  $\Lambda_3$. The corresponding symmetric points have for symmetry group $G^{\mathcal{V}_0}(\mathds{Z})$ the Weyl group of $D_5$ for $D_5$ and $D_5^*$ and the Weyl group of $A_5$ for all the others. In both cases the action of  $G^{\mathcal{V}_0}(\mathds{Z})$ on the tangent space admits exactly two independent invariant quadratic polynomials. It is therefore sufficient to study an automorphic function on an appropriate surface including the symmetric points to determine if they are indeed minima  \cite{Bossard:2023bhv}. This led us to introduce the surface parametrised by $r$ and $x$ as
\be
H(r,x) = \begin{pmatrix}
1& x &\hspace{-1.5mm} x &\hspace{-1.5mm}x & x\\
0 & & \hspace{1.5mm}\mathds{1}& &
\end{pmatrix}^\intercal \begin{pmatrix}
r^{-4} & 0\\
0 & rH_{A_{4}^*}
\end{pmatrix}\begin{pmatrix}
1& x &\hspace{-1.5mm} x &\hspace{-1.5mm}x & x\\
0 & & \hspace{1.5mm}\mathds{1}& &
\end{pmatrix}\;,
\label{hypersurface}
\ee
where \footnote{The parametrisation used in \cite{Bossard:2023bhv} corresponds to the inverse matrix, but is equivalent to this one. Note that the Gram matrix of the root lattices $D_5$ and $A_5$ can be chosen as the Cartan matrix by definition, but it is more convenient to choose a Gram matrix with manifest S$_4$ and S$_5$ symmetry, respectively.}
\begin{equation}
H_{A_4^*}=H_{A_4}^{-1}=\frac{1}{5^\frac{3}{4}}\begin{pmatrix}
4 & -1 & -1 & -1 \\
-1 & 4 & -1 & -1 \\
-1 & -1 & 4 & -1 \\
-1 & -1 & -1 & 4 
\end{pmatrix}\;.
\end{equation}
This hypersurface conveniently contains all the symmetric points with an irreducible matrix $H_L$ except $H_{A_5}$, with
\bea H\left(\frac{6^\frac{1}{5}}{5^\frac{1}{4}},-\frac{1}{5}\right) &=& H_{A_5^*}\; ,  \CR
H\left(\frac{2^\frac{2}{5}}{5^\frac{1}{4}},-\frac{2}{5}\right)&=&H_{D_5^*} \; , \qquad H\left(\frac{1}{2^{\frac{2}{5}} 5^\frac{1}{4}},-\frac{2}{5}\right) = H_{D_5} \CR
H\left(\frac{\left(\frac{3}{2}\right)^\frac{1}{5}}{5^\frac{1}{4}},-\frac{2}{5}\right)&=&H_{A_5^{+2}} \; , \quad \,\, H\left(\frac{\left(\frac{2}{3}\right)^\frac{1}{5}}{5^\frac{1}{4}},-\frac{2}{5}\right) = H_{A_5^{+3}}\; . \eea
Note that the Gram matrix of $L$ is only determined up to an $SL(5,\mathds{Z})$ transformation. In particular the matrices $H_L$ we chose satisfy $H_{L^*} = \gamma^\intercal H_{L}^{-1} \gamma$ for a non-trivial $\gamma \in  SL(5,\mathds{Z})$. For example  $H(\scalebox{0.7}{$\frac{(\frac23)^{\frac15}}{5^{\frac14}}$},-\frac{2}{5}) = H(\scalebox{0.7}{$\frac{(\frac32)^{\frac15}}{5^{\frac14}}$},\frac{3}{5})^{-1}$. 

\begin{figure}[h!]
    \centering
    \includegraphics[scale=0.67]{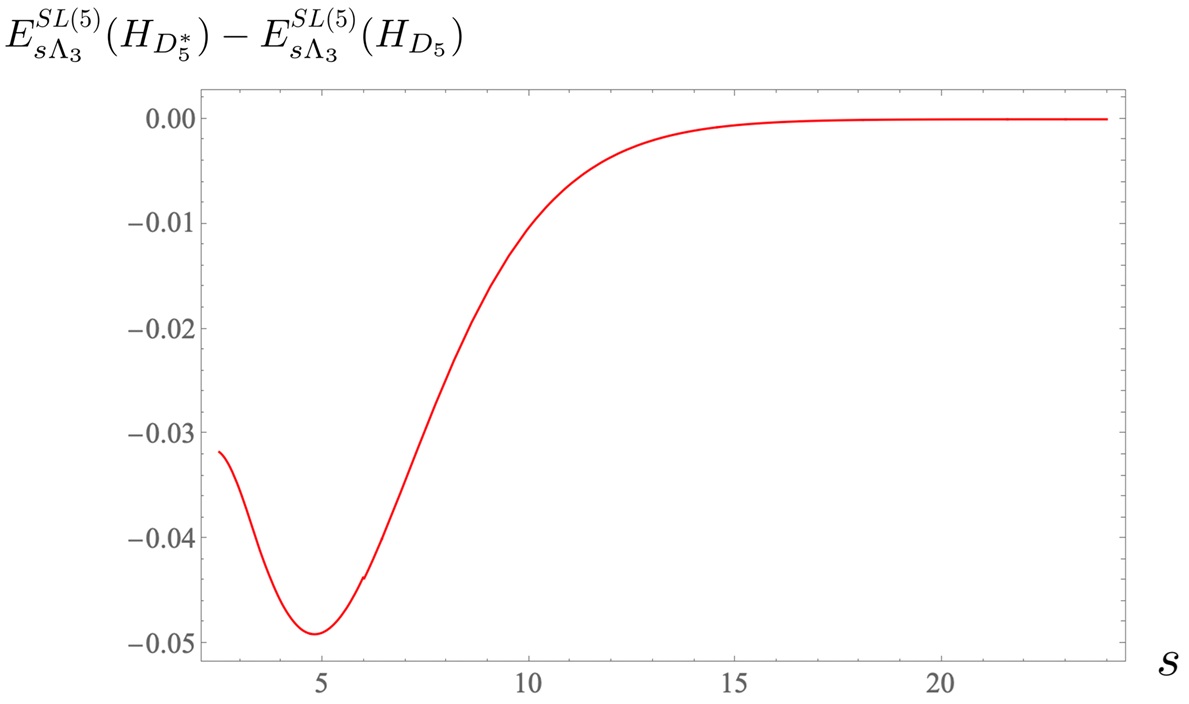}
    \caption{\small The difference between ${E}^{\scalebox{0.55}{$SL(5)$}}_{\scalebox{0.6}{$s\Lambda_{3}$}}$ at $D_5^*$ and $D_5$ is plotted as a function of $s$. For small $s$ the function is plotted using the truncated Fourier expansion \eqref{N-1 Fourier series} discussed above. However since this approximation becomes bad at large $s$ the function must be plotted using the truncated sum \eqref{Eisenstein5 lambda2} where the first $6$ layers of charges $q$ with minimum norm are included for $D_5^*$ and the first $7$ for $D_5$. The two functions are then stitched together at $s=6$. The function changes sign at $s\approx21$.}
    \label{figure1}
\end{figure}

The numerical approximation of $\widehat{E}^{\scalebox{0.55}{$SL(5)$}}_{\scalebox{0.6}{$\frac{5}{2}\Lambda_{3}$}}$ evaluated at $H(r,x)^{-1}$ shows that the function behaves qualitatively as the minimal Eisenstein series  ${E}^{\scalebox{0.55}{$SL(5)$}}_{\scalebox{0.6}{$s \Lambda_{4}$}}$. On the surface parametrised by $H(r,x)$, $\widehat{E}^{\scalebox{0.55}{$SL(5)$}}_{\scalebox{0.6}{$\frac{5}{2}\Lambda_{3}$}}$ has a local minimum at the four symmetric points   $H_{L}$, for $L = D_5,D_5^*,A_5$ and $A_5^*$, of which, only $D_5$ and $D_5^*$ are minima of $\widehat{E}^{\scalebox{0.55}{$SL(5)$}}_{\scalebox{0.6}{$\frac{5}{2}\Lambda_{3}$}}$ in $SO(5) \backslash SL(5)$. The value of $\widehat{E}^{\scalebox{0.55}{$SL(5)$}}_{\scalebox{0.6}{$\frac{5}{2}\Lambda_{3}$}}$ at $D_5$ and $D_5^*$ is almost degenerate, the lowest value being at $D_5^*$ with 
\be
\widehat{E}^{\scalebox{0.65}{$SL(5)$}}_{\frac52\Lambda_3}(H_{D_5^*})\approx68.8311\;.
\ee
The global minimum of $\widehat{E}^{\scalebox{0.55}{$SL(5)$}}_{\scalebox{0.6}{$\frac{5}{2}\Lambda_{1}$}}$ is instead at $L=D_5$. One might guess already that the global minimum of the Wilson coefficient \eqref{Wilson coeff} must be either at $D_5$ or at $D_5^*$, and we shall  indeed find below that it is at $D_5$. In order to prove that the only two local minima of  $\widehat{E}^{\scalebox{0.55}{$SL(5)$}}_{\scalebox{0.6}{$\frac{5}{2}\Lambda_{3}$}}$ are indeed at $D_5$ and $D_5^*$, we would need to systematically study the next-to-minimal Eisenstein series on the entire moduli space. A first step would be to analyse the function on symmetric domains $\Sigma_{\mathcal{V}_0}$ of increasing dimension. Our main argument to trust that the minima must be on the symmetric points is that $\widehat{E}^{\scalebox{0.55}{$SL(5)$}}_{\scalebox{0.6}{$\frac{5}{2}\Lambda_{3}$}}$ behaves qualitatively as the minimal Eisenstein series $\widehat{E}^{\scalebox{0.55}{$SL(5)$}}_{\scalebox{0.6}{$\frac{5}{2}\Lambda_{4}$}}$, and so is expected to have minima at very special geometric points. More generally, one gets that the behaviour of ${E}^{\scalebox{0.55}{$SL(5)$}}_{\scalebox{0.6}{$s\Lambda_{3}$}}$ is qualitatively similar to the one of ${E}^{\scalebox{0.55}{$SL(5)$}}_{\scalebox{0.6}{$s \Lambda_{1}$}}$, except that the minimum value of ${E}^{\scalebox{0.55}{$SL(5)$}}_{\scalebox{0.6}{$s\Lambda_{3}$}}$  is exchanged between $H_{D_5}$ and $H_{D_5^*}$ for different values of $s>\frac{5}{4}$. One finds indeed that the global minimum of ${E}^{\scalebox{0.55}{$SL(5)$}}_{\scalebox{0.6}{$s\Lambda_{3}$}}$ is at $H_{D_5}$ for $s \gtrsim 21 $, and at $H_{D_5^*}$  for $\frac{5}{2}\leq s\lesssim21$ (see Fig. \ref{figure1}), whereas the global minimum of ${E}^{\scalebox{0.55}{$SL(5)$}}_{\scalebox{0.6}{$s \Lambda_{1}$}}$ is at $H_{D_5}$ for all $s>\frac{5}{4}$. 

\vskip 5mm
After symmetric points, the natural loci to look at are the symmetric lines, i.e. sets of points $\Sigma_{\mathcal{V}_0}$ defined as in \eqref{symmetric stratum} which are one-dimensional. We show in Appendix \ref{appendix A} that there is a unique symmetric line invariant under the symmetric group S$_5$ that joins the four symmetric points $H_L$ with $L=D_5,A_5^{+2},A_5^{+3},D_5^*$, and that can be parametrised in the surface as $H(r,- \frac25)$. The plot of the Wilson coefficient on this line
\begin{figure}[h!]
    \centering
    \includegraphics[scale=0.9]{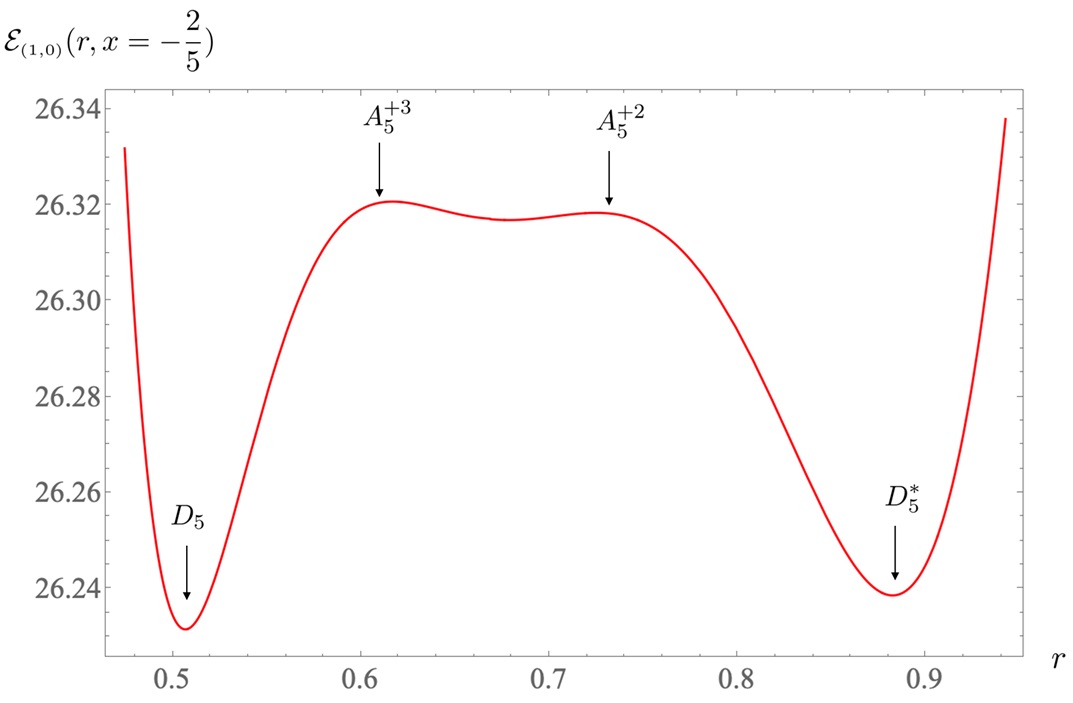}
    \caption{\small Pullback of $\mathcal{E}_\gra{1}{0}$ along the two dimensional $(r,x)$ surface defined by \eqref{hypersurface} plotted on the symmetric line $x=-\frac25$ containing $H_{D_5}$, $H_{A_5^{+3}}$, $H_{A_5^{+2}}$ and $H_{D_5^*}$. Only $H_{D_5^*}$ and $H_{D_5}$ are local minima and $H_{D_5}$ is the global minimum. Because the numerical approximation is bad when $r$ is too small, this plot is obtained by stitching together a plot of \scalebox{0.8}{$\zeta(5)\widehat{E}^{\scalebox{0.65}{$SL(5)$}}_{\frac52\Lambda_1}(r,x=-\frac25)+\frac{\pi}{15}\zeta(5)\widehat{E}^{\scalebox{0.65}{$SL(5)$}}_{\frac52\Lambda_3}(r,x=-\frac25)$} for $r>1/5^{\frac14}$ and a plot of the dual function \scalebox{0.8}{$\zeta(5)\widehat{E}^{\scalebox{0.65}{$SL(5)$}}_{\frac52\Lambda_4}(1/(5^\frac12 r),x=-\frac25)+\frac{\pi}{15}\zeta(5)\widehat{E}^{\scalebox{0.65}{$SL(5)$}}_{\frac52\Lambda_2}(1/(5^\frac12 r),x=-\frac25)$} for $r<1/5^{\frac14}$.}
    \label{figure2}
\end{figure}
shows that the minimum is at $L=D_5$ (see Fig. \ref{figure2}). The plots of $\widehat{E}^{\scalebox{0.65}{$SL(5)$}}_{\frac52\Lambda_1}$ and $\widehat{E}^{\scalebox{0.65}{$SL(5)$}}_{\frac52\Lambda_3}$ are qualitatively similar, with the global minimum being at $D_5^*$ in the second case.\\
 We computed the minimum value as 
\be
 \mathcal{E}_\gra{1}{0}(H_{D_5}) = \zeta(5)\widehat{E}^{\scalebox{0.65}{$SL(5)$}}_{\frac52\Lambda_1}(H_{D_5})+\frac{\pi}{15}\zeta(5)\widehat{E}^{\scalebox{0.65}{$SL(5)$}}_{\frac52\Lambda_3}(H_{D_5})\approx26.2315\;,
\label{minimum value}
\ee
where the displayed digits are stable upon increasing the numerical approximation.

\section{The superstring two-loop integral}\label{The superstring two-loop integral}
In order to fix the  ambiguity in the definition of the Wilson coefficient  associated to the logarithmic divergence of the two-loop integral in supergravity, we need to define a renormalisation scheme using the finite two-loop integral in superstring theory. The two-loop superstring four-graviton amplitude in dimension $D=10-d$ was derived in \cite{DHoker:2005vch}. It is defined as an integral over the moduli space of Riemann surfaces of genus two parametrised by the period matrix  $\Omega^{ij}=\Omega^{ij}_1 + i \Omega^{ij}_2$ defined by the cycle integrals 
\be \oint_{A_i} \omega^j(z) = \delta_i^j \; , \qquad \Omega^{ij} = \oint_{B^i} \omega^j(z) \; , \ee
where $\omega^i$ are the abelian holomorphic one-forms, $A_i$ and $B^i$ define a symplectic basis of cycles and $\Omega_2$ is a positive symmetric matrix. The period matrices 
 \be \Omega= \left( \begin{array}{cc} \rho & \ v  \\  v  &\, \ \sigma  \end{array}\right) \label{PeriodM} 
  \;  \ee are identified in moduli space up to global diffeomorphisms represented by $Sp(4,\mathds{Z})$ matrices $\gamma = (^A_C{}^B_D)$ acting as 
\be \Omega \rightarrow ( A \Omega + B ) ( C \Omega + D)^{-1} \; . \ee
We write $G(z_a,z_b)$ the Green function, and 
\be \Gamma^{\scalebox{0.6}{2-loop}}_{I\hspace{-0.5mm}I_{d,d}}  = \det \Omega_2^{\frac{d}{2}} \sum_{m , n \in \mathds{Z}^d} e^{- \pi  \Omega_2^{ij} \bigl[ G^{IJ} ( m_{i I} + B_{IK} n^{K}_i )  ( m_{j L} + B_{JL} n^{L}_j )  + G_{IJ} n^I_i n^J_j \bigr] + 2\pi i \Omega_1^{ij} m_{i I} n^I_j } \; , \ee
the Narain partition function. Using definition \eqref{ScalarFunction} for the amplitude, one can then write the scalar factor as  \cite{DHoker:2005vch}
\be \mathcal{A}^{\scalebox{0.6}{2-loop}} =  \frac{\pi \alpha^{\prime 3}}{128} \int_{\mathcal{F}_2}  \frac{d^6 \Omega}{(\det \Omega_2)^5}  \Gamma^{\scalebox{0.6}{2-loop}}_{I\hspace{-0.5mm}I_{d,d}}  \int_{\Sigma^4}  \mathcal{Y}_S\wedge \overline{ \mathcal{Y}_S}  \exp\biggl(  - \frac{\alpha^\prime}{2} \sum_{a>b} k_a \!\cdot \! k_b \, G(z_a,z_b)\biggr) \ee
where ${\mathcal{Y}_S}$ is the holomorphic  4-form on four copies of the Riemann genus two surface $\Sigma$ 
\be   \mathcal{Y}_S   = \frac{\alpha^\prime}{3}\bigl( (t-u)\;  \varepsilon_{ij} \varepsilon_{kl} + (s-t) \varepsilon_{ik} \varepsilon_{lj} + (u-s) \varepsilon_{il} \varepsilon_{jk} \bigr)  \omega^i(z_1) \omega^j (z_2)  \omega^k(z_3) \omega^l (z_4)  \; , \label{YS}  \ee
and $\mathcal{F}_2=U(2) \backslash Sp(4,\mathds{R}) / P\hspace{-0.2mm}Sp(4,\mathds{Z})$ is the fundamental domain of $Sp(4,\mathds{Z})$ in the Siegel upper half plane \cite{zbMATH03144647}
\be -1/2 \le \rho_1,v_1,\sigma_1 \le 1/2 \;, \qquad 0 \le  2 v_2 \le  \rho_2\le  \sigma_2  \; , \qquad |\!\det( C \Omega + D)|\ge 1\; ,   \ee
for all $C,D$ such that $\gamma = (^A_C{}^B_D) \in Sp(4,\mathds{Z})$. 

In order to obtain the leading contribution at low energy, it is convenient to decompose the fundamental domain in three regions as was proposed in \cite{Pioline:2015nfa}
\be \mathcal{F}_{2} = \mathcal{F}_{2,\Lambda} \cup \mathcal{F}^\prime_{2,\Lambda,\Lambda_1} \cup \mathcal{F}^{\prime\prime}_{2, \Lambda_1} \;  , \ee
for $\Lambda > \Lambda_1\gg 1$.
The regions are more easily described using the parametrisation
 \be \Omega= \left( \begin{array}{cc} \rho & u_1 {+}  \rho u_2  \\  u_1 {+}  \rho u_2  &\, \sigma_1 {+} i  (L {+} \rho_2 u_2^{\; 2})  \end{array}\right) \label{PeriodMcusp1} 
  \; , \ee
with $L> \rho_2(1-u_2^{\, 2})$ and $0\le u_2\le 1$ in $\mathcal{F}_2$. The three domains are defined  such that 
\bea \mathcal{F}_{2,\Lambda} &=& \mathcal{F}_{2} \cap \{ L \le \Lambda \} \; , \CR
\mathcal{F}^\prime_{2,\Lambda,\Lambda_1}  &=& \mathcal{F}_{2} \cap \{ L\ge \Lambda \ge \Lambda_1 \ge {\rm Im}[\rho] \} \; , \CR
 \mathcal{F}^{\prime\prime}_{2, \Lambda_1} &=& \mathcal{F}_2  \cap \{ L\ge \Lambda, {\rm Im}[\rho]\ge \Lambda_1 \} \; . \eea
The integral over the first domain $ \mathcal{F}_{2,\Lambda}$ is analytic in the Madelstam variables (in the domain of analyticity of the field theory amplitude in the $(s,t)$ plane), and one can expand the integrand in $\alpha'$ in order to define the two-loop contribution to the Wilson coefficient. This gives the contribution to the Wilsonian effective action obtained from the contributions of massive string states running in the two loops. The integral over the second domain  $\mathcal{F}^\prime_{2,\Lambda,\Lambda_1}$ corresponds to the handle of length $L$ of the genus 2 Riemann surface being very long, i.e. the non-separating single degeneration limit. In this degeneration limit one can interpret $L$ as the Schwinger parameter of a one-loop supergravity integral with the insertion of a local operator coming from the one-loop contribution to the Wilsonian effective action. The integral over the third region $ \mathcal{F}^{\prime\prime}_{2, \Lambda_1}$ is restricted to the maximal degeneration limit when the two handles are very large, which gives rise to the supergravity amplitude including $\alpha'$ corrections coming from the insertion of the Wilson coefficients associated to the tree-level string theory effective action  \cite{Tourkine:2013rda}.

The precise shapes of the boundaries between the three domains make the computation difficult. It is therefore very useful to include an additional regularisation through the insertion of $(\det\Omega_2)^\epsilon L^\delta$  in the integral
\be \mathcal{A}_{\epsilon,\delta}^{\scalebox{0.6}{2-loop}} =  \frac{\pi \alpha^{\prime 3}}{128} \int_{\mathcal{F}_2}  \frac{d^6 \Omega}{(\det \Omega_2)^5}  (\det\Omega_2)^\epsilon L^\delta \Gamma^{\scalebox{0.6}{2-loop}}_{I\hspace{-0.5mm}I_{d,d}}  \int_{\Sigma^4}  \mathcal{Y}_S\wedge \overline{ \mathcal{Y}_S}  \exp\biggl(  - \frac{\alpha^\prime}{2} \sum_{a>b} k_a \!\cdot \! k_b \, G(z_a,z_b)\biggr) \; , \ee
for complex parameters $\epsilon$ and $\delta$. For generic complex momenta $k_a$ (such that no loop momenta can be on-shell) the integral converges for arbitrary complex values of $\epsilon$ and $\delta$, and so do the integrals over the three different domains described above. In this paper we are only interested in the leading contribution in $\alpha'$, that contributes to the next-to-leading Wilson coefficient as
\be  \mathcal{E}^{\scalebox{0.6}{2-loop}}_{\gra{1}{0}\, {\Lambda,\epsilon,\delta}}(\varphi) = 4\pi \int_{\mathcal{F}_{2, \Lambda} }   \frac{d^6 \Omega}{(\det \Omega_2)^3}  (\det\Omega_2)^\epsilon L^\delta \Gamma^{\scalebox{0.6}{2-loop}}_{I\hspace{-0.5mm}I_{d,d}}   \; . \label{Wilson2loop}  \ee
This integral converges in the limit $L\rightarrow \infty$ if ${\rm Re}[\epsilon] < 2  - \frac{d}{2}$ and ${\rm Re}[\epsilon+\delta] < 1  - \frac{d}{2}$.

For the integral over the second domain $\mathcal{F}^\prime_{2,\Lambda,\Lambda_1}$ one can take the approximation of the Green function in the non-separating degeneration limit and extract at  leading order in $\alpha'$ \footnote{This limit has not been derived directly using the genus-two Green function in the non-separating degeneration limit, but can be deduced by consistency with the Ward identities and the one-loop amplitude in string theory.}
\bea \mathcal{A}^{\scalebox{0.6}{2-loop}\, \prime }_{\Lambda,\Lambda_1, \epsilon,\delta} \sim  \frac{ \pi \alpha^{\prime 5} s^2}{4} \int_\Lambda^\infty   \hspace{-0.8mm}  \frac{dL}{L^{3-\frac{d}{2}-\epsilon}} \hspace{0.2mm} \int_0^1 \hspace{-0.8mm}  dy_2  \hspace{0.2mm} \int_0^{y_1}\hspace{-1.4mm}  dy_1  \hspace{0.2mm} e^{\alpha' s \pi L y_1(y_2-y_1)} \int_{\mathcal{F}_{1, \Lambda_1}} \hspace{-1.6mm} \frac{d^2 \rho}{\rho_2^{\, 2}} \rho_2^{\epsilon + \delta} \Gamma^{\scalebox{0.6}{1-loop}}_{I\hspace{-0.5mm}I_{d,d}}  +\circlearrowleft \; . \quad \label{Non-Separa} \eea
This integral converges for ${\rm Re}[\epsilon] > 2  - \frac{d}{2}$ and ${\rm Re}[\epsilon+\delta] < 1  - \frac{d}{2}$. This contribution is subleading in seven dimensions, i.e. for $d=3$, so we shall not analyse this term. In general one needs to introduce the two parameters $\epsilon$ and $\delta$ in order to be able to remove the cutoff in the three regions, but because we can disregard the integral \eqref{Non-Separa} over the second domain it is consistent to set $\delta = 0 $. This will simplify the analysis of the integral in the third region. 

The integral over the third domain $ \mathcal{F}^{\prime\prime}_{2, \Lambda_1}$ reduces at leading order to the  2-loop supergravity amplitude. After unfolding the integration domain $P\hspace{-0.1mm}GL(2,\mathds{Z})\backslash GL(2,\mathds{R}) / SO(2)$ one can write the integral over the Schwinger parameter space
\bea  \hspace{-15mm}\mathcal{A}^{\scalebox{0.6}{2-loop} \, \prime\prime}_{> \Lambda_1,  \epsilon}  &\sim&   \frac{\pi \alpha^{\prime 5} }{4} \hspace{-8mm} \int\limits_{\substack{L_I \ge 0\\ \sum_{I<J} \! L_I L_J  \ge \Lambda \Lambda_1\\ 
L_I{+}L_J \ge \Lambda_1}} \hspace{-6mm}  \frac{dL_1dL_2dL_3}{\bigl( \sum_{I<J} \! L_I L_J\bigr)^{5- \frac{d+2\epsilon}{2}}} s^2 \int_{0\le y_1\le y_2 \le 1} \hspace{-12mm} dy_1 dy_2  \; L_1^{\; 2} \; e^{\pi \alpha^\prime  s  L_1 y_1(1-y_2)   }  \label{SgraTwoLoop} \\*
&& \times  \Biggl(\int_{0\le y_3\le y_4 \le 1} \hspace{-12mm} dy_3 dy_4   \; L_2^{\; 2} \; e^{\pi \alpha^\prime    \Bigl( s L_2 y_3(1-y_4)   + \frac{ L_1 L_2 L_3}{\scalebox{0.5}{$\sum\limits_{I<J} L_I L_J$}}  \bigl( t ( y_2-y_1)(y_4-y_3) + s ( 1-y_1-y_4)(1-y_2-y_3)  \bigr)  \Bigr) } \CR
&& +\int_{0}^1 \! dy_3 \int_0^1 \! dy_4   \;  L_2 L_3  \; e^{ \pi \alpha^\prime \frac{L_1 L_2 L_3}{\scalebox{0.5}{$\sum\limits_{I<J} L_I L_J$}}   \bigl( t ( y_2-y_1)(y_4-y_3) + s( 1-y_1-y_4)(1-y_2-y_3)   \bigr)}  \Biggr) +{\rm perm.} \nonumber \eea
 which converges for ${\rm Re}[\epsilon] > \frac{3}{2}  - \frac{d}{2}$.
 
For generic $\epsilon$ the two integrals  \eqref{Wilson2loop} and \eqref{SgraTwoLoop} give formal power series in the cutoff parameters for which the powers of $\Lambda$ and $\Lambda_1$ cancel precisely in the sum by definition. Because \eqref{Wilson2loop} is convergent for ${\rm Re}[\epsilon] < 1 - \frac{d}{2}$ and \eqref{SgraTwoLoop} is convergent for ${\rm Re}[\epsilon] >\frac{3}{2}  - \frac{d}{2}$, one can eliminate the cutoff parameters $\Lambda$ and $\Lambda_1$  and define the string amplitude as the sum of the two meromorphic analytic continuations in $\epsilon$ in the limit $\epsilon \rightarrow 0$. Note that in order to compute the next order in $\alpha'$, or consider the leading contribution in $D\le 6$, one would also need to analyse the integral \eqref{Non-Separa} and therefore  keep both $\epsilon$ and $\delta$ in the three  integrals. But for the present analysis we can safely set $\delta = 0$ from the beginning.   

 The simplification $\delta = 0$ allows to interpret the integral over the third domain as the supergravity two-loop integral in dimensional regularisation in $D=7-2\epsilon$ dimensions \cite{Bern:1998ug}
\bea 
\mathcal{A}^{\scalebox{0.6}{2-loop} \, \prime\prime}_{  \epsilon}&\sim& 16 ( 2\pi)^{2D-6} \alpha^{\prime D-2} \int \frac{d^{D}p d^{D} q}{(2\pi)^{2D}}  \Biggl(  \frac{s^2}{p^2 (p{-}k_1)^2 (p{-}k_1{-}k_2)^2(p{+}q)^2 q^2(q{-}k_4)^2 (q{-}k_3{-}k_4)^2} \CR
&& \hspace{24mm} + \frac{s^2}{p^2 (p{-}k_1)^2(p{-}k_1{-}k_2)^2 (p{+}q)^2 (p{+}q{+}k_3)^2 q^2(q{-}k_4)^2}+{\rm perm.}   \Biggr)   \;  .  \eea

To determine the Wilson coefficient it is useful to use the unfolding formula in the limit in which one circle of the torus $T^3$ of radius $R$ becomes large. To fix conventions we define 
\be G^{-1} = \left(\begin{array}{cc} \frac{1}{R^2} &  \frac{a^\intercal_1}{R^2}  \\  \frac{a_1}{R^2} & G^{-1}_{d{-}1} +\frac{a_1 a_1^\intercal }{R^2}\end{array}\right) \; , \qquad B  = \left(\begin{array}{cc} 0 & a_2^\intercal \\ - a_2  & B_{d{-}1} \end{array}\right) \; , \ee
where $G_{d{-}1}$ and $B_{d{-}1}$ are the metric and $B$-field of the torus $T^d$. One computes using the standard unfolding formula \cite{MR993311,McClain:1986id,Dixon:1990pc,Pioline:2014bra,Florakis:2016boz} that
\bea  \mathcal{E}^{\scalebox{0.6}{2-loop}}_{\gra{1}{0}\, {\epsilon}}(\varphi) &=& 4\pi \int_{\mathcal{F}_{2} }   \frac{d^6 \Omega}{(\det \Omega_2)^3}  (\det\Omega_2)^\epsilon  \Gamma^{\scalebox{0.6}{2-loop}}_{I\hspace{-0.5mm}I_{d,d}}   \\
&=& 4\pi R^2  \int_{\mathcal{F}_{2} }   \frac{d^6 \Omega}{(\det \Omega_2)^3}  (\det\Omega_2)^\epsilon  \Gamma^{\scalebox{0.6}{2-loop}}_{I\hspace{-0.5mm}I_{d{-}1,d{-}1}} \CR
&& + R^2 \int_0^\infty \hspace{-2mm}\frac{dt}{t^{\frac{7-d}{2}}} \int_0^1\hspace{-2mm} du_1\int_0^1\hspace{-2mm} du_2\int_0^1 \hspace{-2mm}d\sigma_1 \int_{\mathcal{F}_1} \hspace{-1mm}\frac{d^2\rho}{\rho_2^{\, 2}} \mathcal{R}_\epsilon(\Omega)  \sum_{k=1}^\infty e^{- \frac{\pi R^2 k^2}{t}} \CR
&& \times \rho_2^{\frac{d-1}{2}} \hspace{-2mm} \sum_{q,p\in I\hspace{-0.5mm}I_{d{-}1,d{-}1}} e^{ \pi i  \rho ( p + u_2 q)_L^2 - \pi i  \bar \rho ( p + u_2 q)_R^2 - \pi t ( q_L^2 + q_R^2) + \pi i  \sigma_1 (q,q) + 2\pi i u_1 (q,p) + 2\pi i k (q,a)} \nonumber \eea
where $\mathcal{R}_\epsilon(\Omega)$ is determined as a modular invariant continuous, but not differentiable, function. On the copy of the fundamental domain obtained from $\mathcal{F}_2$ by the action of $\gamma$, the function $\mathcal{R}_\epsilon(\Omega)$ is defined as 
\be \mathcal{R}_\epsilon(\Omega) = \left( \frac{\det \Omega_2}{|\! \det (C \Omega + D)|^2}\right)^\epsilon \; , \quad \Omega \in \gamma \mathcal{F}_2 \; . \ee
For fixed $t$ and $u_2$, there is always a lower bound for $\rho_2$ such that $|\! \det( C \Omega + D)| >1$ for all $C$ and $D$, and 
\be \mathcal{R}_\epsilon(\Omega)=t^\epsilon \rho_2^\epsilon  \; , \qquad {\rm if}\; \rho_2 >\underset{(c,d)}{\rm max}\Biggl(  \frac{1-d^2 t }{(c + d u_2)^2} , \frac{1}{t} , 1 \Biggr) \; . \ee
This ensures that $\rho_2^{\, \epsilon}$ can be used as a regulator of the $\rho$ integral over the $SL(2,\mathds{Z})$ fundamental domain $\mathcal{F}_1$. The integral converges for ${\rm Re}[\epsilon]<  \frac{2-d}{2}$, and after Poisson summation the expression converges for ${\rm Re}[\epsilon]<  \frac{3-d}{2}$. In the unfolded integral the regulator function $\mathcal{R}_\epsilon(\Omega)$ is only needed in the fundamental domain in which it is equal to $(\det \Omega_2)^\epsilon$. It follows that the difference of the integral with the inclusion of $\mathcal{R}_\epsilon(\Omega)$ and with the inclusion of $(\det \Omega_2)^\epsilon$ is convergent for all $\epsilon$ and therefore analytic in $\epsilon = 0$ where it must then vanish. Up to terms of order $\epsilon$, one can therefore safely assume the integral to be \footnote{Where we set $\epsilon$ to $0$ in the last line, the difference being absorbed in the $\mathcal{O}(\epsilon)$ terms. }
\bea  \mathcal{E}^{\scalebox{0.6}{2-loop}}_{\gra{1}{0}\, {\epsilon}}(\varphi)    &=& 4\pi R^2  \int_{\mathcal{F}_{2} }   \frac{d^6 \Omega}{(\det \Omega_2)^3}  (\det\Omega_2)^\epsilon  \Gamma^{\scalebox{0.6}{2-loop}}_{I\hspace{-0.5mm}I_{d{-}1,d{-}1}}  + \mathcal{O}(\epsilon) \CR
&& + R^2 \int_0^\infty \hspace{-2mm}\frac{dt}{t^{\frac{7-d-2\epsilon}{2}}} \int_0^1\hspace{-2mm} du_1\int_0^1\hspace{-2mm} du_2\int_0^1 \hspace{-2mm}d\sigma_1 \int_{\mathcal{F}_1} \hspace{-1mm}\frac{d^2\rho}{\rho_2^{\, 2}}  \sum_{k=1}^\infty e^{- \frac{\pi R^2 k^2}{t}} \CR
&& \times \rho_2^{\frac{d-1}{2}+\epsilon} \hspace{-2mm} \sum_{q,p\in I\hspace{-0.5mm}I_{d{-}1,d{-}1}} e^{ \pi i  \rho ( p + u_2 q)_L^2 - \pi i  \bar \rho ( p + u_2 q)_R^2 - \pi t ( q_L^2 + q_R^2) + \pi i  \sigma_1 (q,q) + 2\pi i u_1 (q,p) + 2\pi i k (q,a)} \CR
&=&  4\pi R^2  \int_{\mathcal{F}_{2} }   \frac{d^6 \Omega}{(\det \Omega_2)^3}  (\det\Omega_2)^\epsilon  \Gamma^{\scalebox{0.6}{2-loop}}_{I\hspace{-0.5mm}I_{d{-}1,d{-}1}}  + \mathcal{O}(\epsilon) \\
&& + 4\pi \xi(5-d-2\epsilon)  R^{d-3+2\epsilon}    \int_{\mathcal{F}_1} \hspace{-1mm}\frac{d^2\rho}{\rho_2^{\, 2}} \rho_2^{\, \epsilon} \, \Gamma^{\scalebox{0.6}{1-loop}}_{I\hspace{-0.5mm}I_{d{-}1,d{-}1}}\CR
 && + 16 \pi \xi(4-d) R^{\frac{d-1}{2}} \hspace{-3mm} \sum^\prime_{q\in I\hspace{-0.5mm}I_{d{-}1,d{-}1}} \hspace{-2mm} {\rm gcd}(q) \sigma_{d-5}(q) E^{D_{d{-}2}}_{(d{-}4)\Lambda_1}(v_q) \frac{ K_{\frac{d-5}{2}} (2\pi R \scalebox{0.8}{$\sqrt{2 q_R^{\, 2}}$})}{\scalebox{0.8}{$\sqrt{2 q_R^{\, 2}}$}^{\frac{d-3}{2}}} e^{2\pi i (q,a)} \nonumber \; . \eea
The one-loop integral in the next to last line also needs to be computed to extract the pole in $\frac{1}{\epsilon}$. Defining in the same way 
\be G^{-1}_{d{-}1} = \left(\begin{array}{cc} \frac{1}{\tilde{R}^2} &  \frac{\tilde{a}^\intercal_1}{\tilde{R}^2}  \\  \frac{\tilde{a}_1}{\tilde{R}^2} & G^{-1}_{d{-}2} +\frac{\tilde{a}_1 \tilde{a}_1^\intercal }{\tilde{R}^2}\end{array}\right) \; , \qquad B_{d-1}  = \left(\begin{array}{cc} 0 & \tilde{a}_2^\intercal \\ - \tilde{a}_2  & B_{d{-}2} \end{array}\right) \; , \ee
one obtains through the same argument that 
\bea  \int_{\mathcal{F}_1} \hspace{-1mm}\frac{d^2\rho}{\rho_2^{\, 2}} \rho_2^{\, \epsilon} \, \Gamma^{\scalebox{0.6}{1-loop}}_{I\hspace{-0.5mm}I_{d{-}1,d{-}1}}  &=& \tilde{R} \int_{\mathcal{F}_1} \hspace{-1mm}\frac{d^2\rho}{\rho_2^{\, 2}} \rho_2^{\, \epsilon} \, \Gamma^{\scalebox{0.6}{1-loop}}_{I\hspace{-0.5mm}I_{d{-}2,d{-}2}} + 2 \xi(4-d-2\epsilon) \tilde{R}^{d-3+2\epsilon} \CR
&& + 4 \tilde{R}^{\frac{d-2}{2}}  \hspace{-3mm} \sum^\prime_{q\in I\hspace{-0.5mm}I_{d{-}2,d{-}2}} \hspace{-2mm} \sigma_{d-4}(q) \frac{ K_{\frac{d-4}{2}} (2\pi \tilde{R} \scalebox{0.8}{$\sqrt{2 q_R^{\, 2}}$})}{\scalebox{0.8}{$\sqrt{2 q_R^{\, 2}}$}^{\frac{d-4}{2}}} e^{2\pi i (q,\tilde{a})}  + \mathcal{O}(\epsilon) \; . 
 \eea
From this we obtain that the  regularised Wilson coefficient in seven dimensions is the sum of a term analytic at $\epsilon=0$ and the divergent term
\be  \mathcal{E}^{\scalebox{0.6}{2-loop}}_{\gra{1}{0}\, {\epsilon}}(\varphi)  \sim  8\pi \xi(2-2\epsilon)   \xi(1-2\epsilon)  R^{2\epsilon}  \tilde{R}^{2\epsilon}  \; . \ee
This pole cancels the pole of the 2-loop supergravity amplitude in dimensional regularisation, and the sum is finite by construction in the superstring amplitude. To extract the non-perturbative Wilson coefficient one needs to go to Einstein  frame, and therefore move the logarithm of the string coupling constant from the non-analytic amplitude to the Wilson coefficient. The non-analytic term in the string coupling comes from the leading log using \eqref{EinsteinString} \cite{Green:2010sp}
\be - \frac{4\pi^2}{3} {\rm ln} ( \alpha' s ) = - \frac{4\pi^2}{3} {\rm ln} ( \lP^2 s ) + \frac{16\pi^2}{15} {\rm ln}  g_{\scalebox{0.6}{$7$}}\;.  \ee
In dimensional regularisation, this term comes from $\frac{2\pi^2}{3\epsilon} ( \alpha' s)^{-2\epsilon}$ and moving the log of the string coupling constant to the Wilson coefficient can conveniently be done using  
\be 8 \pi \xi(2-2\epsilon) \xi(1-2\epsilon) R^{2\epsilon} \tilde{R}^{2\epsilon}  + \frac{2\pi^2}{3\epsilon}  g_{\scalebox{0.6}{$7$}}^{ \frac{8}{5} \epsilon } = 8 \pi \xi(2-2\epsilon) \xi(2\epsilon) g_{\scalebox{0.6}{$7$}}^{- \frac{8}{5} \epsilon } R^{2\epsilon} \tilde{R}^{2\epsilon}  + \frac{2\pi^2}{3\epsilon}   + \mathcal{O}(\epsilon)\; . \ee
Matching the non-perturbative Wilsonian coefficient, one obtains the complete amplitude contribution as 
\bea 
\mathcal{A}_{  \epsilon}&=& 16  \lP^{10-4\epsilon} \int \frac{d^{7-2\epsilon}p d^{7-2\epsilon} q}{(2\pi)^{6}}  \Biggl(  \frac{s^2}{p^2 (p{-}k_1)^2 (p{-}k_1{-}k_2)^2(p{+}q)^2 q^2(q{-}k_4)^2 (q{-}k_3{-}k_4)^2} \CR
&& \hspace{36mm} + \frac{s^2}{p^2 (p{-}k_1)^2(p{-}k_1{-}k_2)^2 (p{+}q)^2 (p{+}q{+}k_3)^2 q^2(q{-}k_4)^2}+{\rm perm.}   \Biggr)  \CR
&& + 8 \pi  \lP^{10} \frac{s^2+t^2+u^2}{16} \xi(2-2\epsilon) \xi(2\epsilon) E^{\scalebox{0.65}{$SL(5)$}}_{\epsilon \Lambda_2 + \epsilon \Lambda_4} \;  ,  \eea
where 
\bea &&  \xi(2-2\epsilon) \xi(2\epsilon) E^{\scalebox{0.65}{$SL(5)$}}_{\epsilon \Lambda_2 + \epsilon \Lambda_4}  \CR
&=& \xi(2-2\epsilon) \xi(2\epsilon) + \xi(2-2\epsilon) \xi(5-2\epsilon) \Biggl( E^{\scalebox{0.65}{$SL(5)$}}_{(\frac{5}{2}-\epsilon)\Lambda_1} - \frac{\xi(1-2\epsilon)}{\xi(5-2\epsilon)} \Biggr) \CR
&& +\xi(5-2\epsilon)\xi(4-2\epsilon) \Biggl( E^{\scalebox{0.65}{$SL(5)$}}_{(\frac{5}{2}-\epsilon)\Lambda_3} - \frac{\xi(1-2\epsilon)\xi(2-2\epsilon)}{\xi(5-2\epsilon)\xi(4-2\epsilon)} \Biggr)  + \mathcal{O}(\epsilon) \CR
&=& \xi(2-2\epsilon) \xi(2\epsilon) + \xi(2) \xi(5)  \widehat{E}^{\scalebox{0.65}{$SL(5)$}}_{\frac{5}{2}\Lambda_1} + \xi(5)\xi(4) \widehat{E}^{\scalebox{0.65}{$SL(5)$}}_{\frac{5}{2}\Lambda_3}   + \mathcal{O}(\epsilon) \; . \eea

\subsection*{Acknowledgements}

We would like to thank Andrea Guerrieri, Aditya Hebbar, Julio Parra-Martinez and Piotr Tourkine for discussions. 

\appendix

\section{Symmetric surfaces}\label{appendix A}
In this appendix we describe the symmetric lines joining symmetric points inside a symmetric hypersurface $\Sigma_{\mathcal{V}_0}$ \eqref{symmetric stratum} of dimension 3 stabilised by S$_4$, parametrised by 
\be H_{{\rm S}_4}(a,b,c) = (2-c)^{- \frac{3}{5}} b^{- \frac15} \left(\begin{array}{ccccc} 2\, &\, c\, &\, c\, &\, c\, &\, a\\ 
c \, &\, 2\, &\, c\, &\, c\, &\, a\\ 
c\, &\, c\, &\, 2\, &\, c\, &\, a\\ 
c\, &\, c\, &\, c\, &\, 2\, &\, a\\
a\, &\, a\, &\, a\, &\, a\, &\frac{b +4a^2}{2+3c}\end{array}\right) \; . \ee
It contains  representatives of all symmetric points invariant under a maximal finite subgroup, with for example 
\bea D_5 \, &:& \, H_{{\rm S}_4}(2,4,1)\; , \qquad A_5^{+3} \, : \, H_{{\rm S}_4}(2,\tfrac32,1) \; , \qquad A_5^{+2} \, : \, H_{{\rm S}_4}(2,\tfrac23,1)\; ,  \qquad D_5^* \, : \, H_{{\rm S}_4}(2,\tfrac14,1)  \CR
A_5 \, &:& \, H_{{\rm S}_4}(1,6,1) \; , \qquad A_5^* \, : \, H_{{\rm S}_4}(- \tfrac25,\tfrac{24}{25},- \tfrac25)\; . \eea
 In this hypersurface there are symmetric lines invariant under a subgroup S$_5$, one linking the four symmetric points in the first line above through $H_{{\rm S}_4}(2,b,1)$ and one liking $A_5 $ and $A_5^*$ through $H_{{\rm S}_4}(c,(2-c)(4c+2),c)$. Note that there are infinitely many representatives of each symmetric point and each symmetric line in this hypersurface because it is preserved by an infinite modular group, but we only need to consider one representative.

We find that they are the only symmetric lines linking symmetric points invariant under a maximal finite subgroup of $SL(5,\mathds{Z})$. To check this we have computed the intersection subgroup associated to pairs of  symmetric points. We have computed the subgroups homomorphic to $W(D_5)$ that leave invariant the bilinear forms $H_{{\rm S}_4}(2,4,1)$  and $H_{{\rm S}_4}(2,\frac14,1)$, and checked the intersection of the former with an $SL(5,\mathds{Z})$ conjugate of the latter for a large number of matrices   $\gamma \in SL(5,\mathds{Z})$ (more than 100000).\footnote{We have generated the $SL(5,\mathds{Z})$ by multiplying elements of $W(D_5)$ and $W(A_5)$ stabilising the bilinear forms above.} The intersection subgroup is at most S$_5$, can be S$_4$ or a smaller subgroup. We have checked that all the intersection subgroups homomorphic to S$_5$ are conjugate under $W(D_5)$ to the S$_5$ subgroup that stabilises $H_{{\rm S}_4}(2,b,1)$
\be \underset{\gamma \in SL(5,\mathds{Z})}{\rm max} W(D_5,H_{{\rm S}_4}(2,4,1) ) \cap \gamma^{-1} W(D_5^*,H_{{\rm S}_4}(2,\tfrac14,1) ) \gamma \cong {\rm S}_5 \; . \ee
We find the same result for all pairs of symmetric points among $D_5$, $A_5^{+3}$, $A_5^{+2}$ and $D_5^*$. On the contrary, we find no symmetric line linking these points to $A_5$ and $A_5^*$ and the intersection of the two associated $W(A_5)$ is at most S$_5$, can be  S$_3$ or a smaller subgroup. All the intersection subgroups homomorphic to S$_5$ are conjugate under $W(A_5)$ to the S$_5$ subgroup that stabilises $H_{{\rm S}_4}(c,(2-c)(4c+2),c)$.

%\bibliography{biblio}
\bibliographystyle{utphys}
\providecommand{\href}[2]{#2}\begingroup\raggedright\endgroup

\end{document}